\documentclass[12pt,preprintnumbers]{article}
\usepackage[utf8]{inputenc}

\usepackage{amsmath, amssymb, wasysym, slashed, multirow,hyperref}
\usepackage{pdfpages}
\usepackage{cite}
\usepackage{dsfont}
\usepackage{times}
\usepackage[export]{adjustbox}

\usepackage{slashed}

\newcommand{\bitem}{\begin{itemize}}
\newcommand{\eitem}{\end{itemize}}
\newcommand{\bwt}{\begin{widetext}}
\newcommand{\ewt}{\end{widetext}}
\newcommand{\beq}{\begin{equation}}
\newcommand{\eeq}{\end{equation}}

\newcommand{\bdm}{\begin{displaymath}}
\newcommand{\edm}{\end{displaymath}}
\newcommand{\bea}{\begin{eqnarray}}
\newcommand{\eea}{\end{eqnarray}}

\newenvironment{sciabstract}{%
\begin{quote} }
{\end{quote}}

\newcounter{lastnote}

\usepackage{fancyhdr}
\fancypagestyle{plain}{%
	\fancyhead[R]{RBI-ThPhys-2021-15,
ACFI-T21-02}

}

\title{Resurgence of the QCD Adler function }

\author
{Alessio Maiezza,$^{1\ast}$ Juan Carlos Vasquez$^{2\dagger}$\\
\\
\normalsize{$^{1}$Ruder Bo\v skovi\'c Institute, Bijeni\v cka cesta 54, 10000, Zagreb, Croatia,}\\ \\
\normalsize{$^{2}$Amherst Center for Fundamental Interactions, Department of Physics,}\\
\normalsize{University of Massachusetts, Amherst, MA 01003, USA.}\\
\\
\small{ E-mail: amaiezza@irb.hr$^{\ast}$,jvasquezcarm@umass.edu$^{\dagger}$}
}

\date{}

\begin{document}


\baselineskip16pt 


\maketitle


\begin{sciabstract}
We study the QCD Adler function in the energy region $\approx 0.7-2.5$ GeV, in which the non-perturbative effects become dominant.
Our analysis is a renormalon-based evaluation using transseries within the resurgence of the Renormalization-Group-Equation and
does not require the Operator-Product-Expansion.
\end{sciabstract}

\section{Introduction}

Perturbation theory is still the best available tool to deal with Quantum Chromodynamics (QCD) at high energies, where the theory is weakly coupled due to asymptotic freedom~\cite{PhysRevLett.30.1343,PhysRevLett.30.1346}. However, at low energy, perturbation theory fails because of the infrared (IR) divergences and the so-called renormalons~\cite{PhysRevD.10.3235,LAUTRUP1977109,tHooft:1977xjm}. Traditionally, these divergences have been interpreted as the necessity to add non-perturbative information for QCD processes, important in the IR limit. An emblematic example is the one of the Adler function~\cite{PhysRevD.10.3714}, which despite not being observable, can be reconstructed from data by using dispersion relations~\cite{Eidelman:1995ny,Eidelman:1998vc}. The Adler function plays an important role in particle physics, especially in the $e^+e^-$ annihilation into hadrons and inclusive $\tau$ lepton decay -- see Ref.~\cite{Pich:2020gzz} for a recent comprehensive review.

On top of numerical simulations, such as lattice QCD, there are analytical methods to deal with QCD in the IR limit, which use perturbative information together with an integral representation~\cite{Shirkov:1997wi,Nesterenko:2007fm,Cvetic:2008bn}. Other approaches are based on the Operator-Product-Expansion (OPE)~\cite{wilson1972,Parisi:1978iq} in which power corrections, motivated by renormalon singularities are considered together with perturbative calculations~\cite{Beneke:1998ui,Cvetic:2018qxs}.

In this work, we also propose a new renormalon-based evaluation of the Adler function in the IR region. However, instead of employing corrections due to non-renormalizable operators, we use resurgent techniques. Resurgence~\cite{Ecalle1993} is a way to perform a unique analytic continuation of a non-Borel-summable function using its asymptotic expansion. In general, given an asymptotic --divergent-- expansion, it is not possible to assign a unique function to it. However, if we consider certain types of asymptotic expansions, which are (for example) solutions of a particular type of differential equation, we can assign this unique function. Moreover, the resurgence theory generalizes the concept of Borel-Laplace resummation, potentially fixing the ambiguities of the Laplace integral, and has been proposed as a unified framework for perturbative and non-perturbative QFT~\cite{Dunne:2012ae}. Related ideas have been developed in the context of resurgence~\cite{Clavier:2019sph,Borinsky:2020vae,Fujimori:2021oqg},  resurgent extrapolations~\cite{Costin:2019xql,Costin:2020hwg}, renormalons for multi-coupling models~\cite{Maiezza:2018pkk}, beyond Gevrey-1 formal series~\cite{Cavalcanti:2020osb} and combinatorics in Feynman diagrams~\cite{Mahmoud:2020vww}.

We shall employ the proposal of Refs.~\cite{Maiezza:2019dht,Bersini:2019axn}, in the specific framework of resurgence for non-linear, ordinary differential equations (ODEs)~\cite{Costin1995,CostinBook}. The resummation of the renormalons is guaranteed to work because Green functions satisfy the Renormalization Group Equation (RGE). For instance, the ultra-violet (UV) renormalons in the $\phi^4$-model can be resummed in terms of a single-parameter transseries~\cite{Maiezza:2019dht}, and such resummation is justified by the underlying ODE derived from the RGE~\cite{Bersini:2019axn}. The concept is further elaborated in Refs.~\cite{Maiezza:2020nbe,Maiezza:2020qib}.
As a result, the infinite ambiguities due to the renormalons are reduced to a single unknown arbitrary constant appearing in the transseries expansion of the solution to the RGE. This constant cannot be calculated due to the lack of a semi-classical limit for renormalons. Nevertheless, it can be matched with data to fix it. We shall do this procedure for the IR renormalons of the Adler function in QCD, which is related to the two-point Green function of two massless quark currents.

\section{Renormalization group equation and renormalons in the two-point correlation function}\label{Sec:main}

We implement the \emph{resurgence of the renormalization group equation} (RRGE), proposed in Refs.~\cite{Maiezza:2019dht,Bersini:2019axn}, for the QCD Adler function. The basic idea is that the two-point Green function is an analyzable function satisfying the RGE and its renormalon singularities can be resummed~\cite{Maiezza:2019dht}.

Consider the two-point correlation function of two massless quark currents $j_{\mu}= \bar{q}\gamma_{\mu}q$
\begin{equation}\label{twopointfunction}
-\mathrm{i} \int d^{4} x \mathrm{e}^{-i q x}\left\langle 0\left|T\left(j_{\mu}(x) j_{v}(0)\right)\right| 0\right\rangle=\left(q_{\mu} q_{v}-q^{2} g_{\mu v}\right) \Pi\left(Q^{2}\right)\,,
\end{equation}
where $Q^2=-q^2$. Defining $L=\ln(\frac{Q^2}{\mu^2})$, the function $\Pi(L, \alpha_s)$ obeys RGE, namely~\footnote{As shown in Refs.~\cite{Weeks:1980qu,Beneke:2006mk,Collins:2005nj}, in the $\overline{MS}$-scheme the conserved current $j_{\mu}$ has a non-zero anomalous dimension.}
\begin{equation}\label{RGE}
\left[-\partial_{L}+\beta(\alpha_s) \partial_{\alpha_s}- \gamma(\alpha_s)\right] \Pi(L, \alpha_s)=0\,,
\end{equation}
where $\beta(\alpha_s)=\mu^2\frac{d\alpha_s}{d\mu^2}= \beta_0\alpha_s^2+\beta_1\alpha_s^3+\mathcal{O}(\alpha_s)^4$. Without any loss of generality one can define
\begin{equation}\label{definition}
\Pi(\alpha_s) : = \sum_{i=0}^{\infty} \gamma_i(\alpha_s)L^i + R(\alpha_s).
\end{equation}
The series represents the perturbative expansion and
the function $\gamma_0(\alpha_s)$ is the perturbative finite part, which does not vanish when $L\rightarrow 0$;
the function $R(\alpha_s)$ is a non-perturbative correction to the two-point Green function. The expression~\eqref{definition} was also suggested in Refs.~\cite{DeRafael:1974iv,Broadhurst:1992si}.

Notice that in perturbation theory $\gamma_0(\alpha_s)$ is finite and hence the renormalization condition $\gamma_0=1$ is well defined. This does not hold at the non-perturbative level, and in particular, it does not hold for the renormalon contributions, which give a $\gamma_0\sim n!$ and hence an ill-defined expression~\footnote{This is nothing but the manifestation of the impossibility of defining the pole mass at the non-perturbative level~\cite{Bigi:1994em,Beneke:1994sw}.}. In order to stress this last point, we have defined Eq.~\eqref{definition} in a slightly different way from the original Ref.~\cite{Bersini:2019axn}.

From Eq.~\eqref{RGE}, it is straightforward to derive the following equation for the function $R(g)$
\begin{equation}\label{main}
\frac{d R(\alpha_s)}{d \alpha_s} = \frac{ q}{\beta_0 \alpha_s^2} R(\alpha_s) +\frac{\beta_0(a_0q+a+s)-\beta_1\,q}{\beta_0^2\,}\frac{R(\alpha_s)}{\alpha_s}
+ a_0\left(\frac{a}{\beta_0}-1	\right) +\mathcal{O}(R(\alpha_s)^2)\,
\end{equation}
where the anomalous dimension $\gamma(\alpha_s) =\gamma_1(\alpha_s) +q\, R(\alpha_s)+\frac{1}{2}(2s \alpha_sR(\alpha_s))+\mathcal{O}(R^2|\alpha_s R) $, its perutbative part $\gamma_1(\alpha_s)= a \alpha_s +\mathcal{O}(\alpha_s^2)$, in agreement with Ref.~\cite{Bersini:2019axn}; finally $\gamma_0(\alpha_s):= 1+a_0\alpha_s+ \mathcal{O}(\alpha_s^2)$.

The Borel transform of the solution $B[R(g)]\equiv \tilde{R}(z)$ to Eq.~\eqref{main} has an infinite number of singularities at $ q/\beta_0,\, q\in \mathbb{N}$. The coefficient of the term $R(\alpha_s)/\alpha_s$, which we shall denote $a_p$, determines the kind of pole in $\tilde{R}(z)$ as follows $(z+ q/\beta_0)^{-1-a_p}$. The non-linearity in $R$, whose explicit form is irrelevant for our purpose, is the source of the infinite number of poles in $\tilde{R}(z)$. Furthermore, the non-linearity in Eq.~\eqref{main} leads also to logarithmic branch points: if one starts with a pole $(z+ q/\beta_0)^{-1-a_p}$, the non-linear term recursively generates a pole $(z+ (q+1)/\beta_0)^{-1-a_p}$ and a $\log(z+q/\beta_0)$~\footnote{See Refs.~\cite{CostinBook,Dorigoni:2014hea} for the analysis of this kind of equations. In particular, $R$ and $\alpha_s$ have to be identified with the expansion variables $y$ and $1/x$ of Ref.~\cite{CostinBook} respectively.}.

The parameters $q$ (and $a_p$) in Eq.~\eqref{main} are not calculable \textit{a priori} but can be found by matching with an explicit computation.
Since the $n-$bubble computation of Ref.~\cite{tHooft:1977xjm} exhibits the same type of singularities of the solution to Eq.\eqref{main}, one must thus identify $R$ with: the Borel-Ecalle resummation (synthesis in Ref.~\cite{Maiezza:2019dht}) of the ultraviolet (UV) renormalons for non-asymptotically free models when $q=1$; the synthesis of the IR renormalons for an asymptotically free model when $q=-1$. In this work, we are interested in the IR case. The parameter $a_p$ depends on the particular model under consideration and will be discussed for QCD in what follows.

In summary, the solution of Eq.~\eqref{main} can be written in terms of a single-parameter transseries~\cite{CostinBook}. As a consequence,
the two-point Green function can be reconstructed up to an arbitrary
constant due to the unknown initial condition of Eq.~\eqref{main}. This new approach reduces the infinite number of arbitrary constants~\footnote{ Which are usually parameterized using the OPE e.g. see Ref.~\cite{Parisi:1978iq}.} stemming from the renormalons to one parameter that we shall denote $C$.

\section{The Adler function and the resurgence of the renormalization group equation }

The Adler function $D$ is
\begin{equation}
D\left(Q^{2}\right)=4 \pi^{2} Q^{2} \frac{\mathrm{d} \Pi\left(Q^{2}\right)}{\mathrm{d} Q^{2}}\,,
\end{equation}
where $\Pi(Q^2)$ is defined in Eq.~\eqref{twopointfunction}. The Adler function can be written at any order in perturbation theory as follows
\begin{equation}
D_{pert}\left(Q^{2}\right)=1+\frac{\alpha_{s}}{\pi} \sum_{n=0}^{\infty} \alpha_{s}^{n}\left[d_{n}\left(-\beta_{0}\right)^{n}+\delta_{n}\right].
\end{equation}
The perturbative expression is known up to $n=3$~\cite{Gorishnii:1990vf,Surguladze:1990tg,Kataev:1995vh} i.e. up to $\mathcal{O}(\alpha_s^4)$ (see table 4 of Ref.~\cite{Beneke:1998ui}). In this paper, we consider the contribution of the so-called ``fermion bubble graphs" to the Adler function. Since the renormalons dominate the large order behavior of the perturbative coefficients $d_n$~\cite{Beneke:1998ui}, they can be used to estimate the leading contributions to the Adler function. This approach is known in the literature as ``Naive non-Abelianization". In practice, we use the full perturbative expression for $D(Q^2)$ up to $\mathcal{O}(\alpha_s^4)$ and, for the higher-order corrections, we assume the fermion bubble diagrams dominate -- i.e. $\delta_n\sim0$ for $n\geq 4$. This procedure allows us to estimate $d_n$ to all orders in perturbation theory. The Borel transform for the fermion-bubble contribution to the Adler function is given by~\cite{Broadhurst:1992si},
\begin{equation}\label{ADLER_SIMPLY}
\frac{1}{K}B[D_{bubble}](u)=\sum_{n=0} \frac{d_{n}}{n !} u^{n}=\frac{32}{3}\left(\frac{Q^{2}}{\mu^{2}} \mathrm{e}^{C}\right)^{-u} \frac{u}{1-(1-u)^{2}} \sum_{k=2}^{\infty} \frac{(-1)^{k} k}{\left(k^{2}-(1-u)^{2}\right)^{2}},
\end{equation}
where $u=-\beta_0\, z$, $C=-5/3$ in the $\overline{\mathrm{MS}}$-scheme and $K$ is an overall, arbitrary constant of the large order behavior. 

\paragraph{Resurgent approach:} Eq.~\eqref{main} dictates the position and the nature of the singularities in the Borel transform. For instance, $a_p=0,1$ correspond to simple and quadratic poles in the Borel transform, respectively. As we shall see, from the direct computation of the fermion-bubble diagram contribution to the Adler function in QCD~\cite{Neubert:1994vb}, one sees an infinite string of quadratic poles starting at $z=-3/\beta_0$, thus we have to set $a_p=1$ in Eq.~\eqref{main} and identify the function $R$ with the synthesis of the quadratic poles. The remaining simple pole at $z=- 2/\beta_0$ cannot be included in the generalized resummation procedure, and its associated ambiguity is then parameterized in terms of one arbitrary constant after the Laplace transform (which we call $c_1$) as follows
\begin{equation}\label{ambiguity}
\left(z+\frac{n+1}{\beta_0}\right)^{-m} \mapsto -2\pi c_n\, \alpha_s^{1-m} e^{\frac{n+1}{\beta_0\alpha_s}}\,.
\end{equation}
The expression in Eq.~\eqref{ADLER_SIMPLY} is not suitable for the RRGE since it would need to be written as a sum over the singularities in the Borel real axis. To this end, we take the original integral representation given by Neubert~\cite{Neubert:1994vb} and rewrite it as~\footnote{We find the UV renormalon contribution to the Adler function negligibly small in the energy range considered and we omit it in Eq.~\eqref{rearrange}.  Although  the first UV renormalon is the closest singularity to the origin, it does not lead to ambiguities in the Laplace transform.   An improvement to perturbation theory was done in Ref.~\cite{Caprini:2020lff}  using the conformal mapping for UV renormalons. }
\begin{align}\label{rearrange}
\frac{1}{C_F\,K}B[&D_{bubble}](z) = \frac{3 e^{10/3} \mu ^4}{2 \beta_0Q^4 \left(\frac{2}{\beta_0}+z\right)}
+ \frac{e^5 \mu ^6 \left(6 \log \left(\frac{\mu
 ^2}{Q^2}\right)+1\right)}{6\beta_0 Q^6
 \left(\frac{3}{\beta_0}+z\right)} -\frac{e^5 \mu^6}{\beta_0^2 Q^6 \left(\frac{3}{\beta_0}+z\right)^2} -\nonumber\\
& \sum_{p=1}^{\infty}\left( \frac{ \mu ^4 e^{\frac{10 p}{3}+\frac{10}{3}}
 \left(\frac{Q}{\mu }\right)^{-4 p} \left(12 p^2 \log
 \left(\frac{\mu ^2}{Q^2}\right)+20 p^2+6 p \log \left(\frac{\mu
 ^2}{Q^2}\right)-2 p-3\right)}{6 \beta_0p^2 (2 p+1)^2 Q^4
 \left(\frac{2 p+2}{\beta_0}+z\right)} + \right. \nonumber \\
 & \frac{ \mu ^6 e^{\frac{10 p}{3}+5} \left(\frac{Q}{\mu
 }\right)^{-4 p} \left(12 p^2 \log \left(\frac{\mu
 ^2}{Q^2}\right)+20 p^2+18 p \log \left(\frac{\mu
 ^2}{Q^2}\right)+18 p+6 \log \left(\frac{\mu
 ^2}{Q^2}\right)+1\right)}{6 \beta_0(p+1)^2 (2 p+1)^2 Q^6
 \left(\frac{2 p+3}{\beta_0}+z\right)}\nonumber \\
 & \left. -\frac{ \mu ^6 e^{\frac{10 p}{3}+5} \left(\frac{Q}{\mu}\right)^{-4 p}}{\beta_0^2 (p+1) (2 p+1) Q^6
 \left(\frac{2 p+3}{\beta_0}+z\right)^2}+\frac{
 \mu ^4 e^{\frac{10 (p+1)}{3}} \left(\frac{Q}{\mu }\right)^{-4
 p}}{\beta_0^2 p (2 p+1) Q^4 \left(\frac{2
 p+2}{\beta_0}+z\right)^2}\right)\,, \nonumber \\
 \end{align}
where $C_F=\frac{4}{3}$. This expression reproduces the IR renormalon structure in Eq.~\eqref{ADLER_SIMPLY} -- choosing the renormalization scale $\mu^2= Q^2e^{-5/3}$. According to the RRGE, the solution of Eq.~\eqref{main} comes from
the leading poles in the Borel transform. Therefore, we have to consider only the quadratic poles and, separately, the first simple pole at $z=\frac{2}{\beta_0}$:
\begin{align}\label{Borel_resurgence}
\frac{1}{K\,C_F}B[&D_{bubble}](z) \rightarrow \frac{3 e^{10/3} \mu ^4}{2 \beta_0Q^4 \left(\frac{2}{\beta_0}+z\right)}
 -\frac{e^5 \mu^6}{\beta_0^2 Q^6 \left(\frac{3}{\beta_0}+z\right)^2} -\nonumber\\
 & \sum_{p=1}^{\infty} \left[ \frac{
 \mu ^4 e^{\frac{10 (p+1)}{3}} \left(\frac{Q}{\mu }\right)^{-4
 p}}{\beta_0^2 p (2 p+1) Q^4 \left(\frac{2
 p+2}{\beta_0}+z\right)^2} -\frac{ \mu ^6 e^{\frac{10 p}{3}+5} \left(\frac{Q}{\mu}\right)^{-4 p}}{\beta_0^2 (p+1) (2 p+1) Q^6
 \left(\frac{2 p+3}{\beta_0}+z\right)^2} \right]\,. \nonumber \\
 \end{align}
\begin{figure}
 \centering
 \includegraphics[width=0.6\columnwidth]{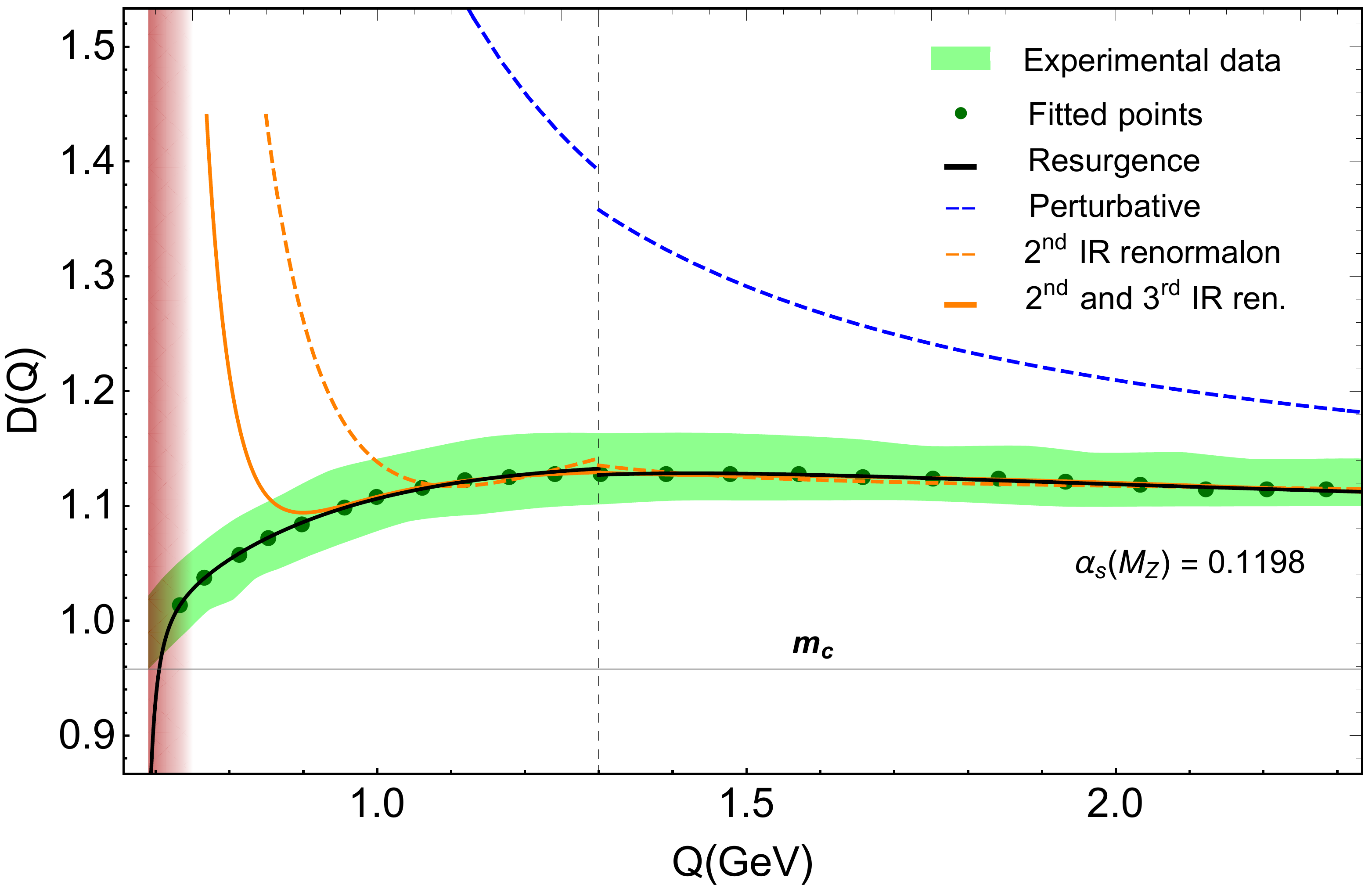}
 \caption{QCD Adler function in the energy range $\sim 0.7-2.5$ GeV: the light-green area is the experimental allowed region; dashed blue line is the perturbative prediction;
 the solid black line is our resurgent result; darker green dots are the points used to fit the constants $C$, $K$ and $c_1$; the dashed orange line represents the fit including only the renormalon at $-2/\beta_0$; the solid orange line represents the fit including only the renormalons at $-2/\beta_0$ and $-3/\beta_0$. Finally, the shaded red band on the left denotes the region where our approximation progressively stops working. }
 \label{fig_Adler}
\end{figure}
In the RRGE approach, the Adler function can be written at leading order as
\begin{equation} \label{trans_adler}
D(Q^2) = D_0(Q^2)-\frac{4\pi}{\beta_0} c_1e^{\frac{2}{\beta_0\,\alpha_s(Q^2)}}+ C e^{\frac{1}{\beta_0\,\alpha_s(Q^2)}}\left(\frac{1}{\alpha_s(Q^2)}\right)^{a_p}D_1(Q^2)\,,
\end{equation}
where $D_0(Q^2)$ contains the perturbative expression plus the higher order $n!$ corrections due to the fermion-bubble diagrams, which are proportional to $K$ and are regularized by taking the principal value of the Laplace integral of Eq.~\eqref{Borel_resurgence}. The constant $a_p =1+ \mathcal{O}(\beta_1/\beta_0^2)$. In what follows, we will neglect the two-loop corrections proportional to $\beta_1$. The function $D_1(Q^2)$ is found from $D_0$~\cite{Maiezza:2019dht} using resurgent relations, and we get from Eq.~\eqref{Borel_resurgence} the following expression for the disconnected part of the Adler function (choosing the renormalization scale $\mu^2= Q^2e^{-5/3}$)
\begin{align}
D_1(Q^2) =& \frac{8 \pi K }{3 \alpha_s \beta_0^2}\left[2e^{\frac{1}{\alpha_s\beta_0}} -
 \left(e^{\frac{1}{\alpha_s\beta_0}}+1\right)
 \log \left(1-e^{\frac{2}{\alpha_s \text{$\beta
 $0}}}\right)\right.-
 &\left. 2 \left(e^{\frac{1}{\alpha_s \text{$\beta
 $0}}}+1\right) \tanh ^{-1}\left(e^{\frac{1}{\alpha_s
 \beta_0}}\right)\right]\,. \label{nonpert_adler}
\end{align}
In the above expression, we substitute the one-loop expression for the running coupling $\alpha_s(Q^2)$, such that $\Lambda_{QCD}^2/Q^2= e^{\frac{1}{\beta_0\alpha_s(Q^2)}}$. We use $\alpha_s(M_Z) = 0.1198$~\cite{Zyla:2020zbs}.

Our strategy is to find both the overall normalization of the large order expression $K$, the non-perturbative constant $C$, and $c_1$ by fitting the ``experimental" values of the Adler function using the Eqs.~\eqref{trans_adler} and~\eqref{nonpert_adler}. We show our \emph{main} result in Fig.~\ref{fig_Adler}: the red shaded area indicates the energy region where the coupling approaches the Landau pole. The dashed orange line is the contribution assuming the IR renormalon at $-2/\beta_0$ dominates the large order behavior; the solid orange line shows the improvement including also the IR renormalon at $-3/\beta_0$ (we name the associated ambiguity as $c_2$, consistent with the parametrization in Eq.~\eqref{ambiguity}). Finally, the black line is our new result using the RRGE.

As can be seen from Fig.~\ref{fig_Adler}, the approximations with one and two IR renormalons drastically improve the perturbative result but they fail below the $m_c$ threshold. Conversely, our improvement fits the experimental values of the Adler function in the range between $E \sim (0.8-2)$ GeV, up to the region in which the coupling approaches the Landau pole. For higher energy, the non-perturbative corrections are exponentially suppressed and perturbation theory is recovered. In particular, it is worth appreciating the difference between the resurgent result (black line) and the two renormalon case (solid orange line) because both require three unknown constants to be fitted ($K,C,c_1$ and $K,c_1,c_2$, respectively). Thus, since both examples have the same number of inputs, the comparison clearly illustrates our improvement.
The value for the $C$, $K$ and $c_1$ from our fit are given by
\begin{equation}
 C, K, c_1 \simeq \left\{
 \begin{array}{ll}
  -0.023,1.41, -0.51\,& \text{ } Q < m_c,\\
  -8.88,0.99,-5.27\,& \text{ } Q \geq m_c.
 \end{array} \right.
 \end{equation}
The different values of $C$ and $K$ at the charm quark mass $m_c$ are due to the discontinuous change in the number of active flavors when the energy reach $m_c\simeq 1.3$ GeV. For completeness, when only the renormalon at $-2/\beta_0$ (dashed orange line) is considered $K,c_1= 0.67, -1.73 $ for $Q<m_c $ and $K,c_1= 0.01, -1.42 $ for $Q\geq m_c $. When only the first two renormalons at $-2/\beta_0$ and $-3/\beta_1$ are considered (solid orange line) $K,c_1,c_2= 1,-6.31,18.3$ for $Q<m_c $ and $K,c_1,c_2= -0.3, -14.3,98.7$ for $Q\geq m_c $.

\paragraph{Remarks.} Let us discuss now possible issues and interpretations of our result within the RRGE.

\begin{itemize}

\item Our approach requires the presence of the first IR renormalon because the structure of the Borel transform of Eq.~\eqref{main} is characterized by poles at $z=- n/\beta_0$, being $n$ an integer between $[1,\infty)$~\cite{Bersini:2019axn}.
 In particular, since the renormalon diagrams~\cite{Neubert:1994vb} gives quadratic poles, the Borel transform of the Adler function should be of the form
\begin{equation}\label{general_structure}
\tilde{D}= \frac{r}{(z-1/\beta_0)^2} + \frac{s}{(z-2/\beta_0)^2} + \frac{t}{(z-3/\beta_0)^2}+...\,.
\end{equation}
However, the first IR renormalon at $z=-1/\beta_0$ does not appear in the explicit calculation of the renormalon diagrams~\cite{tHooft:1977xjm,Neubert:1994vb}, i.e. $r=0$. This is the well-known first renormalon puzzle. As argued in Ref.~~\cite{10.1143/PTPS.131.107}, its presence must be related to UV effects and would then justify why it is not found in the IR expansion done in Ref.~\cite{Neubert:1994vb}. Furthermore, the explicit computation of Ref.~\cite{Neubert:1994vb} gives $s=0$ as well.
We argue that the absence of $s$ is due to the incompleteness of the computation of Ref.~\cite{Neubert:1994vb} since it was done in the large $N_f$ (number of fermions) limit and does not include the gauge boson contributions.
In conclusion and following well-known results, we have set $r,s=0$ in the resummation of the Adler function.

\item The Borel transform of Eq.~\eqref{main} has poles at $(z+ n q/\beta_0)^{-1-a_p}$, which turn out to be quadratic for the specific case considered in this work. However, notice the coefficient $a_p$ receives a contribution proportional to $\beta_1/\beta_0^2$~\cite{Bersini:2019axn} (see also below Eq.~\eqref{main}), in agreement with the known renormalon structure~\cite{Parisi:1978iq,Gardi:2001wg}. This leads to the appearance of multiple non-perturbative sectors corresponding to higher powers of $C$ in Eq.~\eqref{trans_adler}. These higher order terms are expected to be subleading in the energy range under consideration -- below $1\%$ corrections to the leading result. Three and higher loop terms of the beta function do not enter in the transseries~\eqref{trans_adler} since, as emphasized in Refs.~\cite{Maiezza:2019dht,Bersini:2019axn}, only $\beta_0$ and $\beta_1$ are relevant for Eq.~\eqref{main}, making the transseries of Eq.~\eqref{trans_adler} scheme-independent.

\end{itemize}

\section{Summary}

In this work, we propose a renormalon-based approximation of the QCD Adler function using a Borel-Ecalle resummation procedure. We provide an improvement to perturbation theory and as a result, we get a function that accurately follows the behavior of the data up to the Landau pole energy scale  $\simeq 0.7$ GeV. Around this scale, the coupling diverges and the transseries expansion ceases to work.  In contrast  to the operator-product-expansion, there are three unknown constants within this new approach, whereby the model is predictive. Notice that in this sense, it could have been wrong (which by itself is an achievement) and in fact, it starts failing   outside the energy  region of applicability.  Finally, since we have  fixed the constant $C$  in the transseries expansion of the Adler function in Eq.~\eqref{trans_adler}, the corrections to the heavy quark masses of Refs.~\cite{Bigi:1994em,Beneke:1994sw} can be unambiguously computed. It might also be interesting to merge our approach with the analytic one of Refs.~\cite{Shirkov:1997wi,Nesterenko:2007fm,Cvetic:2008bn}. These issues are left for future work.

\section*{Acknowledgement}

AM was partially supported by the Croatian Science Foundation project number 4418. JCV was supported in part under the U.S. Department of Energy contract DE-SC0015376.


\bibliographystyle{jhep}
\bibliography{biblio}

\end{document}